**Static–Dynamic Correlations and Complex Spin-Wave Eigenmodes in Single- and Multilayer Diamond-Shaped Nanomagnets Without Bias Field**

Krishna Begari*

School of Sciences (Physics), National Institute of Technology, Tadepalligudem, Andhra Pradesh 534101, India

Author: Electronic address: krishnabegari.phy@gmail.com

**ABSTRACT**

Diamond-shaped nanomagnets provide a suitable platform for developing microwave devices with reconfigurable characteristics. In this study, the static and dynamic magnetic behaviour of single layer and multilayer diamond-shaped nanomagnets was systematically investigated using micromagnetic simulations. Two distinct remanent magnetic configurations were obtained through a simple magnetic field initialization process. These configurations exhibited different magnetization patterns and dynamic responses. Their resonance characteristics could be modified by applying a nanosecond-scale magnetic field pulse, enabling reconfigurable microwave operation. A clear resonance-frequency shift in the sub-GHz (0.9 GHz) range was observed for the single-layer structure, while the multilayer structure exhibited a significantly larger frequency shift in the GHz range (3 GHz). The larger frequency tunability observed in the multilayer configuration arises primarily from the enhanced dipolar coupling between the vertically coupled magnetic layers, which modifies the local effective magnetic field and consequently influences the magnetization dynamics. These findings demonstrate that diamond-shaped nanomagnets can provide a simple and effective route toward reconfigurable microwave functionality, with potential applications in ultralow-power, ultrafast, and frequency tunable microwave devices.

## I. INTRODUCTION

Microwave magnetic devices are large, and their performance mainly depends on the properties of magnetic materials [1]. In the context of the development of nanotechnology, certain nanostructures have become very important, among which nanopatterns and their arrays are significant [2,3, 4]. Devices operating in the GHz frequency range have gained increasing attention due to their growing importance in recent studies. These nanodevices not only contribute to miniaturization but also play an important role in the development of integrated on-chip microwave systems [5-8]. Compared with conventional electronic and photonic devices, microwave magnetic devices provide greater flexibility because their dynamic microwave response can be modified through a variety of external parameters. This tunability can be realized by applying an external magnetic field to nanodots [9, 10], rings [11], magnonic crystals [12], laser-induced heating [13], the Oersted magnetic field generated by an electric current [14] and the heated tip of a scanning probe microscope [15]. A representative example is provided by magnonic crystals, which consist of periodically arranged magnetic nanostructures. The microwave responses of artificial magnonic crystals and periodic nanostructure arrays can be tuned by modifying structural parameters such as geometry [5,6,16], lattice symmetry [2,4], and magnetic configuration [17,18]. Recent studies on geometrically engineered magnonic systems have further demonstrated the ability to control their microwave behaviour through these parameters [19,20]. Their ferromagnetic resonance (FMR) characteristics can be tailored by modifying the magnetic configuration of the nanostructures or by varying the applied external magnetic field [2,12].

However, in the design of integrated microwave devices at the micron and sub-micron scale, tunability that depends on an external bias magnetic field has become a major limitation. In this context, research has been rapidly expanding on nanomagnetic structures that operate without the need for an external bias magnetic field (bias-field-free) and can control microwave properties based on remanent magnetic states [21]. So far, such bias-field-free microwave reconfigurability has been demonstrated only in a few structures, including dipolar-coupled nanopillars [22,23], nanowires [3,24-26], rhomboid-shaped nanomagnets [27], arrow-shaped nanomagnets [28], and nanomagnetic multilayers [29-31]. These structures exhibit more than one stable remanent magnetic state. Since each remanent state possesses a distinct microwave dynamic response, they can be effectively utilized in the design of reconfigurable microwave devices. At present, next-generation on-chip microwave technology requires nanomagnetic structures with ultra-low power consumption, ultrafast operation, and the ability to be reconfigured without the need for an external bias magnetic field.

Here, single- and multilayer diamond-shaped nanomagnets that exhibit reconfigurable microwave responses are investigated in the absence of an external bias field ($H_{ext} = 0$). In particular, the nanomagnets are arranged in two different ways. One is based on inter-layer spacing, and the other is a multilayer structure separated by a nanomagnetic layer. The magnetization reversal mechanisms and microwave dynamic properties were analysed using the micromagnetic simulation method [32]. This simulation method serves as an effective tool for evaluating the performance of the proposed structures before carrying out the expensive nanofabrication process. Furthermore, the simulated results are consistent with previously reported experimental observations in similar nanomagnetic systems [17]. In addition, fast switching between different remanent magnetic states was achieved, and the distinct microwave dynamic response corresponding to each state was also investigated. The results obtained from this study suggest that these structures can contribute to the design of next-generation nanoscale microwave devices with low power consumption, high operating speed, and tunability without the need for an external bias magnetic field.

## II. METHODS

The magnetic and microwave properties were investigated using static and dynamic micromagnetic simulations with open-source Object Oriented Micromagnetic Framework (OOMMF) software [33]. The OOMMF simulation software is a finite-difference-based solver for the Landau–Lifshitz–Gilbert (LLG) equation, which approximates the spatial derivatives using finite differences. The LLG equation contains precessional and damping terms, as shown below. The spatial domain and time intervals are discretized into finite steps to obtain magnetization as a function of time. In the finite-difference method (FDM), the spatial domain is divided into discrete cells, while the simulation time is advanced in discrete time steps. This discretization enables the magnetization dynamics to be evaluated at successive spatial locations and time intervals. The temporal evolution of the magnetization is governed by the Landau–Lifshitz–Gilbert (LLG) equation,[34] as given below

$$\frac{dM}{dt} = -\gamma M \times H_{eff} + \frac{\alpha}{M_S} M \times \frac{dM}{dt}$$

in normalized magnetization form

$$m = \frac{M}{M_S}$$

$$\frac{dm}{dt} = -\gamma m \times H_{eff} + \alpha m \times \frac{dm}{dt}$$

The magnetization dynamics were studied by numerically solving the Landau–Lifshitz–Gilbert (LLG) equation in its Gilbert damping form. In this equation, $\gamma$ −denotes the gyromagnetic ratio of the electron, whereas $\alpha$ −represents the dimensionless Gilbert damping constant. The effective magnetic field ($H_{\text{eff}}$) includes contributions from the externally applied magnetic field, demagnetizing field, exchange field, anisotropy field, and other magnetic interactions. The first term of the LLG equation describes the precessional motion of the magnetization about the effective magnetic field, while the second term represents the damping mechanism that gradually drives the magnetization toward its equilibrium configuration. Static micromagnetic simulations were employed to evaluate the magnetic properties, whereas dynamic simulations were performed to investigate the microwave response of the proposed nanostructures.

The material parameters of Permalloy ($Ni_{80}Fe_{20}$) used in the micromagnetic simulations were as follows [27,35,36]: Saturation Magnetization ($M_S$)= 800 emu/cm$^3$, Exchange constant ($A$) = 13x10$^{-7}$ erg/cm, Magnetocrystalline anisotropy ($K_1$) = 0 and damping constant ($\alpha$)= 0.008. The sample is discretized into cubic cells, and the length of the cell size should be less than the exchange length of the sample. Micromagnetic simulations of the dipolar-coupled structures were performed using a mesh size of $5 \times 5 \times 5$ nm$^3$, which is sufficiently smaller than the exchange length of Permalloy (5.7 nm). Static micromagnetic simulations were performed to examine the magnetic hysteresis loops and remanent magnetic states, with a damping constant 0.5 employed to accelerate the computational convergence and reduce the simulation time. Magnetization reversal was simulated using magnetic field steps of 20 Oe. The magnetization dynamics were then carried out using damping parameter ($\alpha$) = 0.008. In order to obtain the simulated FMR spectra, a *sinc* pulse field $(H_S) = H_0\, sinc(2\pi f_c \tau)$ was applied along the x-axis. In the sinc pulse equation, the parameters used were ($H_0$ = *50 Oe*), $f_N$ denotes the Nyquist frequency, which is higher than the cut-off frequency ($f_C$), i.e., $f_N > f_C$, while $\tau$ represents the simulation time. The time-dependent magnetization was recorded up to 4 ns with a time step of *10 ps*. The recorded data were then processed using a Python program to generate the FFT spectra and two-dimensional spatial FMR mode profiles were obtained by analyzing the spatial and temporal magnetization data. All data processing and analysis were performed using Python.

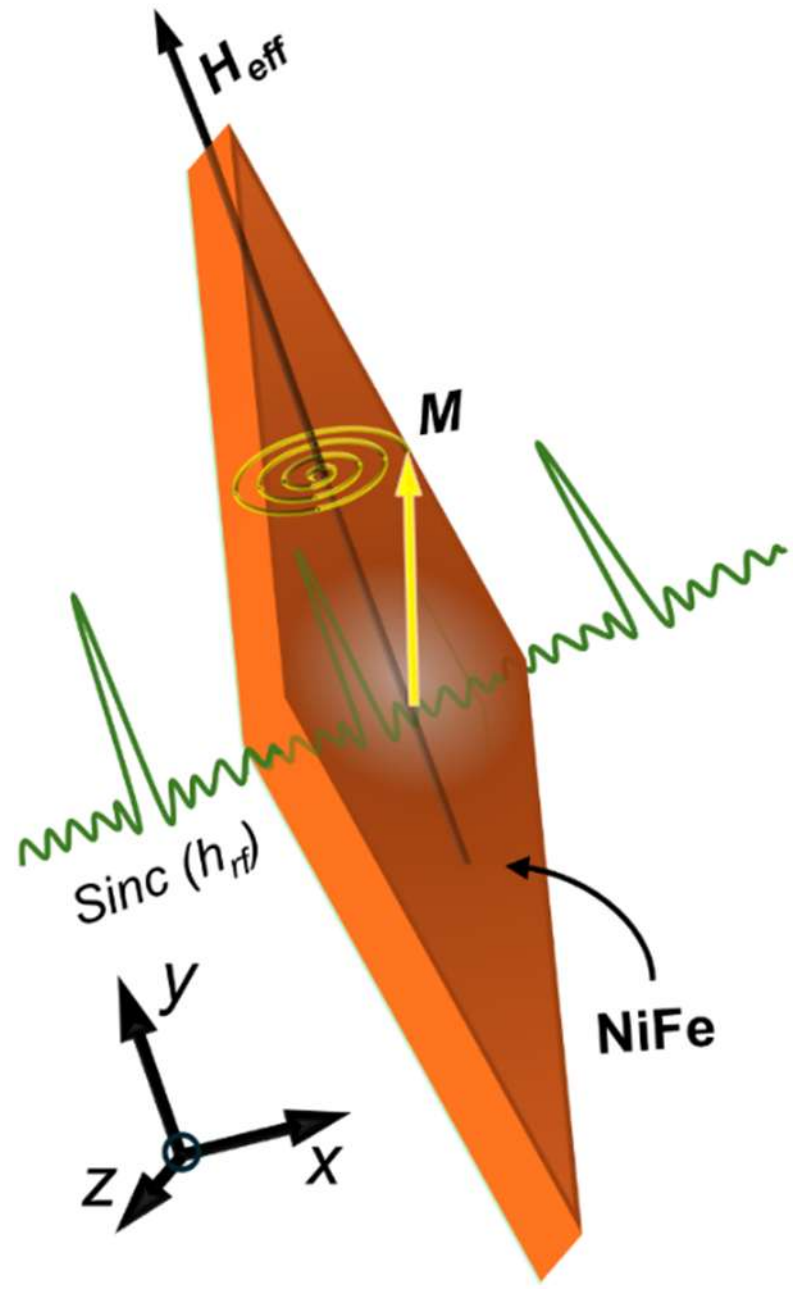

Fig. 1. Schematic representation of the micromagnetic simulation setup for calculating the dynamic response of the Ni-Fe DNM nanomagnet. A *sinc pulse* is applied along the -axis to excite the magnetization dynamics. The magnetization undergoes damped precessional motion about the effective magnetic field, and the resulting time-dependent magnetization is recorded for further analysis.

## III. RESULTS

### 3.1 Static and dynamic properties of a single-layer dipolar coupled nanomagnets

#### 3.1.1. Magnetization reversal and remanent states of SL-DNM

To investigate the magnetic dipolar field behaviour in diamond-shaped nanostructures, five different single-layer configurations, namely SL-1M, SL-2M, SL-2M^, SL-3M, and SL-4M, were designed. These configurations are based on a diamond-shaped nanomagnet (DNM) with dimensions of length ($L$) = 450 nm, width ($w$) = 130 nm, and thickness ($t$) = 25 nm. Here, SL denotes a single-layer structure, while the accompanying number indicates the number of nanomagnets in each configuration. The geometries of these nanostructure configurations are shown in Fig. 2(a–e). Among them, SL-1M consists of a single isolated diamond-shaped nanomagnet, whereas SL-2M, SL-2M^, SL-3M, and SL-4M comprise two, three, or four nanomagnets arranged in different geometrical configurations, resulting in varying strengths of dipolar interaction between neighboring nanomagnets.

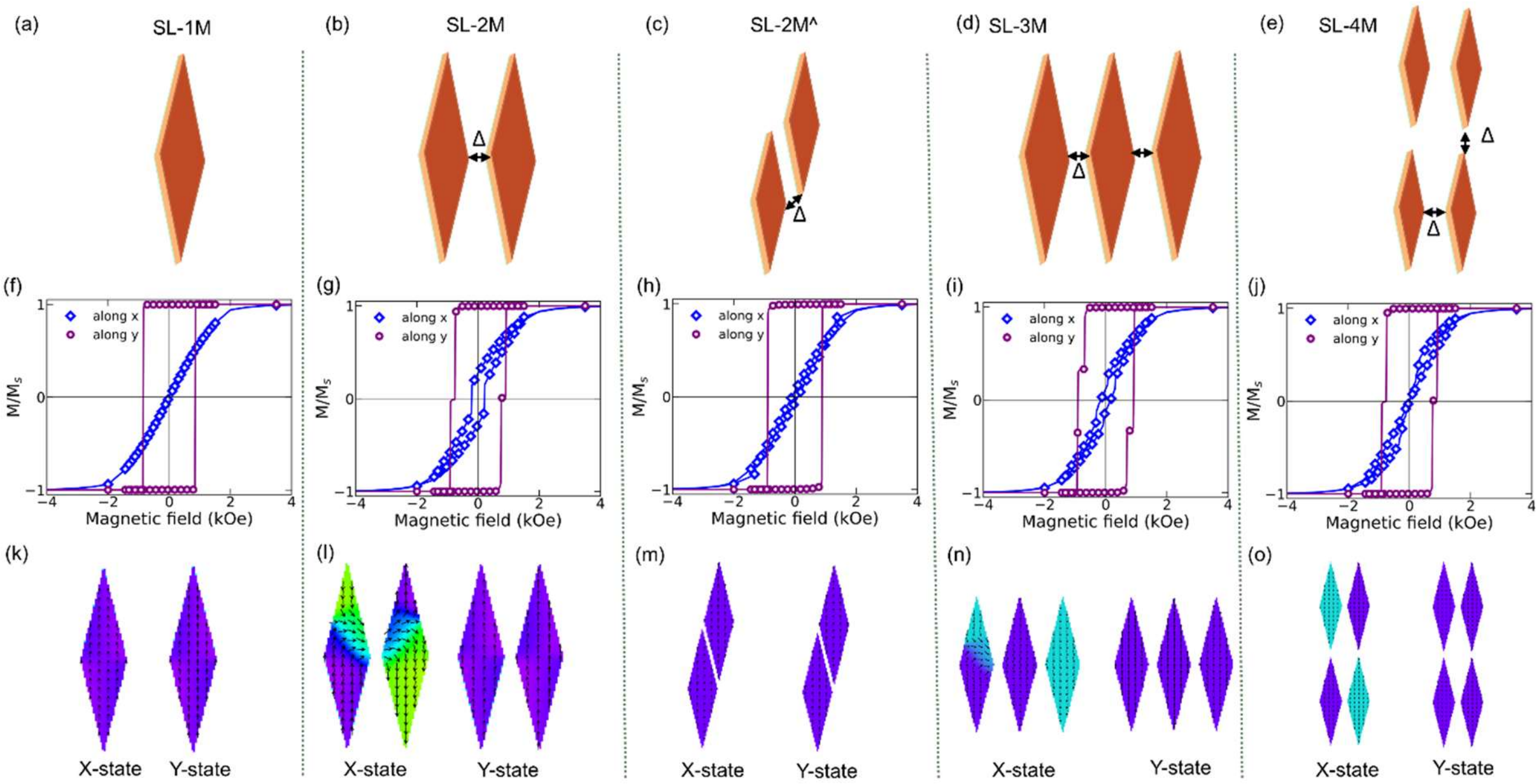

Fig. 2. Static magnetic properties of single-layer diamond-shaped nanomagnets and their networks. (a–e) Schematic illustrations of the nanomagnet configurations. (f–j) Simulated magnetization reversal (hysteresis loops) with the magnetic field applied along the x- and y-directions. (k–o) Corresponding remanent magnetization states (X and Y states) of the nanomagnet systems.

The magnetic reversal characteristics of each configuration were investigated by simulating the hysteresis loops under external magnetic fields applied separately along the x- and y-directions, as shown in Fig. 2(f–j). The hysteresis curves are indicated with symbols '◊' and '○' for the x (short)- and y (long)- axis, respectively. For an isolated nanomagnet (SL-1M), the magnetization changes gradually as the magnetic field is swept along the x-axis, whereas a relatively sharp and single step reversal is observed when the field is applied along the y-axis. As the number of interacting nanomagnets increases, the magnetization reversal process is modified due to stronger magnetostatic interactions between neighboring nanomagnets. In particular, the SL-2M and SL-3M configurations exhibit multiple switching events during reversal along the x-direction, indicating that the nanomagnets switch sequentially rather than simultaneously. On the other hand, SL-2M^ and SL-4M show comparatively smoother reversal behaviour, demonstrating that the magnetization switching process is strongly influenced by the geometrical arrangement of the interacting nanomagnets. Two stable remanent magnetic states were obtained through a magnetic field initialization process. The X-state was achieved by applying a sufficiently large magnetic field along the x-axis followed by reducing the field to zero ($H_x$: 2000 Oe → 0 Oe). Similarly, the Y-state was realized by initializing the structure with the magnetic field applied along the y-axis and subsequently removing it ($H_y$: 2000 Oe → 0 Oe). The resulting equilibrium magnetization configurations for all the structures are shown in Fig. 2 (k–o). The remanent magnetic states depend strongly on the geometrical arrangement of the diamond-shaped nanomagnets, reflecting the combined effects of shape anisotropy and dipolar interactions. Some configurations maintain an almost uniform magnetization after the external field is removed, whereas others exhibit non-uniform magnetization patterns arising from magnetostatic interactions between neighbouring nanomagnets. These remanent states offer multiple stable magnetic configurations without requiring any modification to the physical geometry of the structures.

Overall, the results demonstrate that the magnetization reversal process and the remanent magnetic states can be tuned simply by changing the number and spatial arrangement of the diamond-shaped nanomagnets. This geometrical approach provides an efficient route for controlling the static magnetic behaviour and establishes a promising platform for the development of bias-field-free reconfigurable microwave and magnonic devices. (37,38)

**3.1.2. Ferromagnetic Resonance and Spin-Wave Modes of SL-DNMs (at H = 0)**

The microwave dynamics of the diamond-shaped nanomagnets were evaluated by calculating their FMR spectra at zero external magnetic field ($H_{\text{ext}} = 0$). The FMR spectra corresponding to the X-state and Y-state remanent magnetic configurations of the SL-1M, SL-2M, SL-3M, and SL-4M structures are shown in Fig. 3(a–d). These spectra extend over a frequency range of approximately 5–12 GHz. Multiple resonance peaks are observed depending on the number of interacting nanomagnets and the initialized magnetic state. A clear difference between the X-state and Y-state spectra is observed for all the structures. This indicates that the remanent magnetization configuration formed after field initialization plays an important role in determining the resonance characteristics.

To understand the nature of the resonance peaks, the corresponding two-dimensional (2D) dynamic magnetization distributions were analysed using Python programming and results are presented in Fig. 3(e–h). The extracted mode profiles reveal the spatial characteristics of the resonant modes at different resonance frequencies for each nanomagnet configuration. For the isolated single-layer nanomagnet (SL-1M), four major resonance modes are observed in both the X-state and the Y-state. The lowest-frequency mode exhibits nearly uniform precession throughout the nanomagnet. However, one or more nodal regions appear in the higher-frequency modes, indicating the presence of increasingly complex spin-wave excitations. Since the remanent magnetization configurations in the X-state and the Y-state are nearly identical, no significant change is observed in their corresponding mode profiles. In contrast, the coupled nanomagnet structures (SL-2M, SL-3M, and SL-4M) exhibit more complex resonance behaviour due to the dipolar interactions between neighbouring nanomagnets. The following resonant modes were observed: center mode (c), splitting mode ($c_e$), nodal-line mode (n), and higher-order spin-wave mode (m). In these structures, the dynamic magnetization extends over multiple nanomagnets, resulting in both collective and localized resonance modes. The lower-frequency modes correspond to in-phase oscillations across the coupled nanomagnets. On the other hand, the higher-frequency modes clearly exhibit additional nodal lines and localized oscillations within each nanomagnet. As the number of interacting nanomagnets increases from SL-2M to SL-4M, the spatial complexity of the resonance modes gradually increases because of stronger dipolar coupling and mode hybridization.

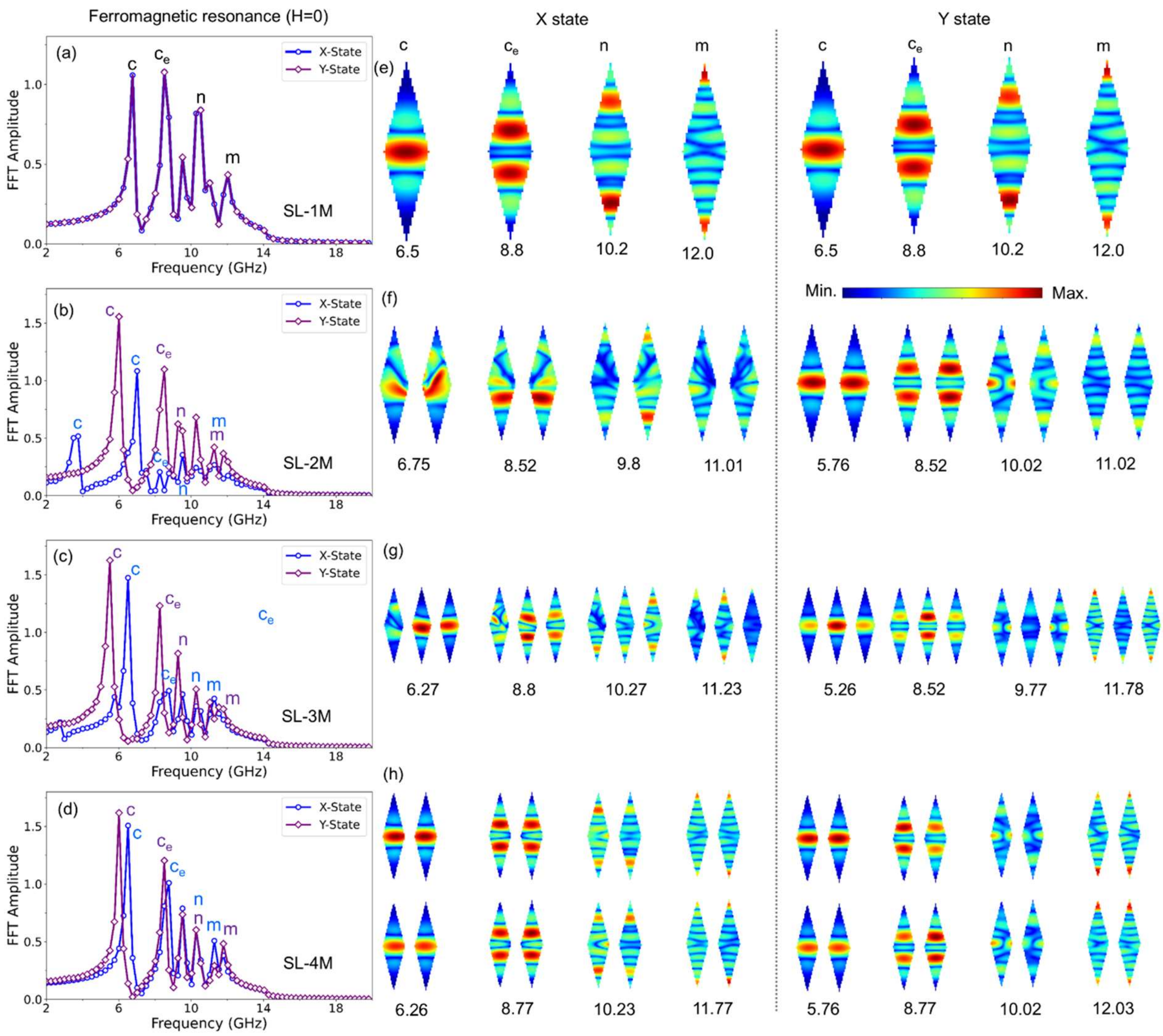

*Fig. 3.* Simulated ferromagnetic resonance (FMR) spectra and corresponding two-dimensional (2D) spatial distributions of the dynamic magnetization for the X and Y remanent states of single layer nanomagnet systems. (a–d) FMR spectra of SL-1M, SL-2M, SL-3M, and SL-4M, respectively. (e–h) Corresponding 2D dynamic mode profiles at the dominant resonance frequencies, illustrating the center (c), center-edge ($c_e$), nodal-line (n), and higher-order spin-wave (m) modes.

A comparison between the X-state and Y-state clearly shows that the spatial distributions of several resonance modes are modified due to the change in the initialization direction. Although some modes retain nearly the same resonance frequency in both remanent states for example modes at 8.2 and 10.2 GHz, others exhibit noticeable frequency shifts along with distinct changes in the localization of the dynamic magnetization. These variations arise from the differences in the internal magnetic field distributions associated with the X-state and Y-state magnetization configurations. The frequency shifts between the X and Y states with respect to the center mode were observed to be 0, 0.99, 1.01, and 0.05 GHz for SL-1M, SL-2M, SL-3M, and SL-4M, respectively.

Overall, the FMR spectra together with the corresponding 2D mode profiles demonstrate that the resonance frequencies as well as the nature of the spin-wave excitations can be effectively controlled by changing the number of diamond-shaped nanomagnets, their geometrical arrangement, and the remanent magnetic states. These characteristics indicate that the proposed structures can serve as an effective platform for the development of bias-field-free reconfigurable microwave and magnonic devices without changing the physical geometry of the nanomagnet structure.

### 3.2 Isolated Multilayer Diamond-Shaped Nanomagnet (ML-DNM)

#### 3.2.1 Distinct Remanent States in an Isolated Multilayer Nanomagnet

To strengthen the effect of dipolar interactions, an isolated DNM (ML-1M) was developed by vertically stacking (multilayer) two Permalloy magnetic layers separated by a thin non-magnetic spacer layer, as shown in Fig. 4(a). In contrast to the single-layer structure, the multilayer configuration exhibits enhanced dipolar coupling between the magnetic layers because their vertical separation is considerably smaller than the lateral spacing between adjacent nanostructures. The multilayer consists of a 25 nm thick bottom magnetic layer ($t_b$), a 15 nm thick non-magnetic spacer, and a 15 nm thick top magnetic layer ($t_t$). The lateral dimensions of the diamond-shaped nanomagnet were kept constant at L = 450 nm and w = 130 nm. The magnetization reversal behaviour of the ML-1M structure was examined by simulating the hysteresis loops with the external magnetic field applied separately along the x- and y-directions, as shown in Fig. 4(b). The hysteresis loop obtained along the x-axis shows a gradual change in magnetization. In contrast, the loop measured along the y-axis exhibits a clear two-step switching process. The intermediate plateau observed in the y-axis loop indicates that the top and bottom magnetic layers switch at different magnetic field values. As a result, an intermediate magnetic configuration is formed before complete magnetization reversal. This switching behaviour arises from the combined influence of shape anisotropy and interlayer dipolar coupling.

Two stable remanent magnetic configurations were obtained by initializing the structure with magnetic fields applied along different directions (2000 to 0 Oe). The corresponding equilibrium magnetization states are presented in Fig. 4(c) and Fig. 4(d). After initialization along the x-axis, the bottom and top magnetic layers settle into antiparallel magnetization configuration, designated as the X-state. In comparison, initialization along the y-axis causes both magnetic layers to remain in parallel magnetization orientations, resulting in the Y-state. The difference between these two remanent states reflects the competition between the dipolar interaction across the layers and the shape anisotropy of each magnetic layer. Because the bottom magnetic layer is thicker than the top layer, it requires a larger magnetic field to switch. In contrast, the thinner top layer switches more easily under the influence of the dipolar field. Consequently, two different and stable remanent magnetic states can be achieved in the isolated multilayer nanomagnet without changing its physical geometry.

These results show that vertical stacking is an effective method for controlling the magnetic configurations of isolated diamond-shaped nanomagnets. Furthermore, the proposed multilayer structure provides a suitable platform for realizing reconfigurable microwave responses without an external bias magnetic field ($H = 0$).

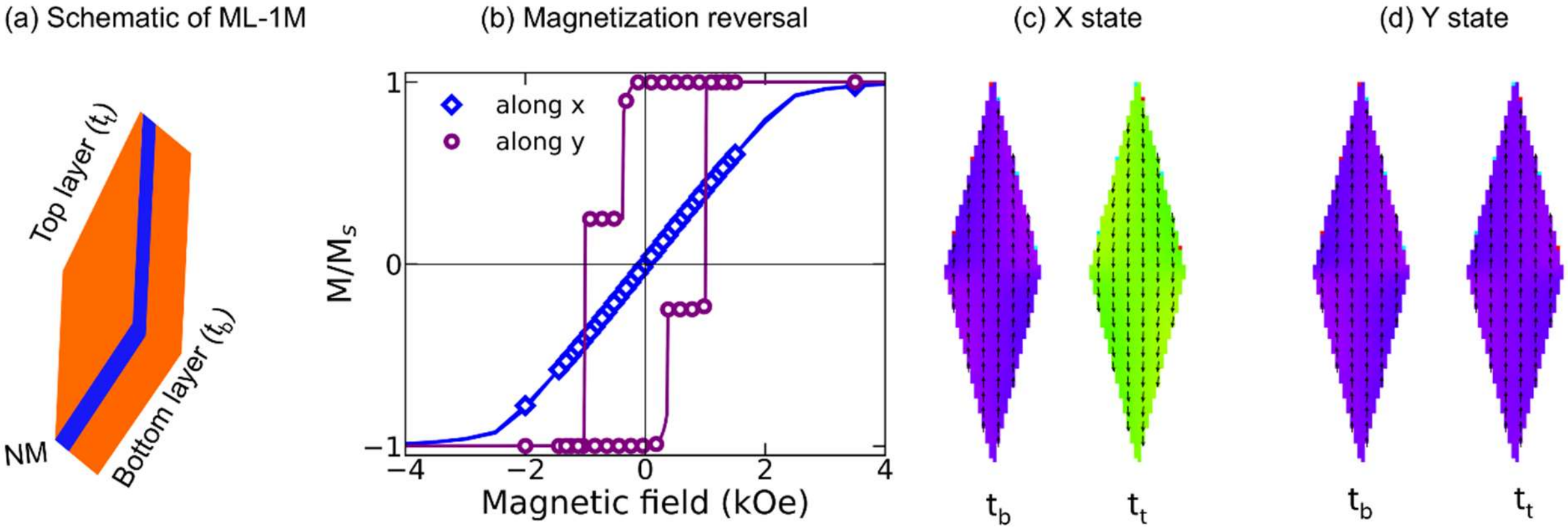

*Fig. 4.* Static magnetic properties of the ML-1M multilayer nanomagnet. (a) Schematic of the multilayer structure. (b) Simulated hysteresis loops measured along the x- and y-directions. (c, d) Remanent magnetization states corresponding to the X and Y configurations in the bottom ($t_b$) and top ($t_t$) magnetic layers

### 3.2.2 Microwave properties in isolated multilayer nanomagnets at $H_{ext} = 0$

The dynamic microwave characteristics of the ML-1M were investigated by calculating the ferromagnetic resonance (FMR) spectra under zero external magnetic field ($H_{ext} = 0$) for two stable remanent magnetic states. The FMR spectra corresponding to the X-state and Y-state are presented in Fig. 5(a). The corresponding two-dimensional (2D) spatial mode profiles are shown in Fig. 5(b) and Fig. 5(c). Multiple resonance modes are observed within the 3–16 GHz frequency range. Their resonance frequencies vary depending on the remanent magnetic state. In addition, a clear frequency shift is observed between the resonance peaks of the X-state and the Y-state. This indicates that the equilibrium magnetization configuration has a significant influence on the dynamic response of the multilayer nanomagnet. To investigate the spatial distribution of the resonance modes, the corresponding 2D spatial profiles were analyzed. Based on the spatial characteristics observed in the mode profiles, four different resonance modes, namely c, $c_e$, n, and m, were identified. The c-mode is mainly concentrated near the central region of the nanomagnet. The $c_e$-mode exhibits magnetization oscillations over a larger portion of the structure. The n-mode is characterized by the presence of nodal regions. The m-mode exhibits a more complex magnetization distribution with additional nodal regions (mixed mode). The mode profiles also show that the spatial distribution of the dynamic magnetization changes as the resonance frequency increases.

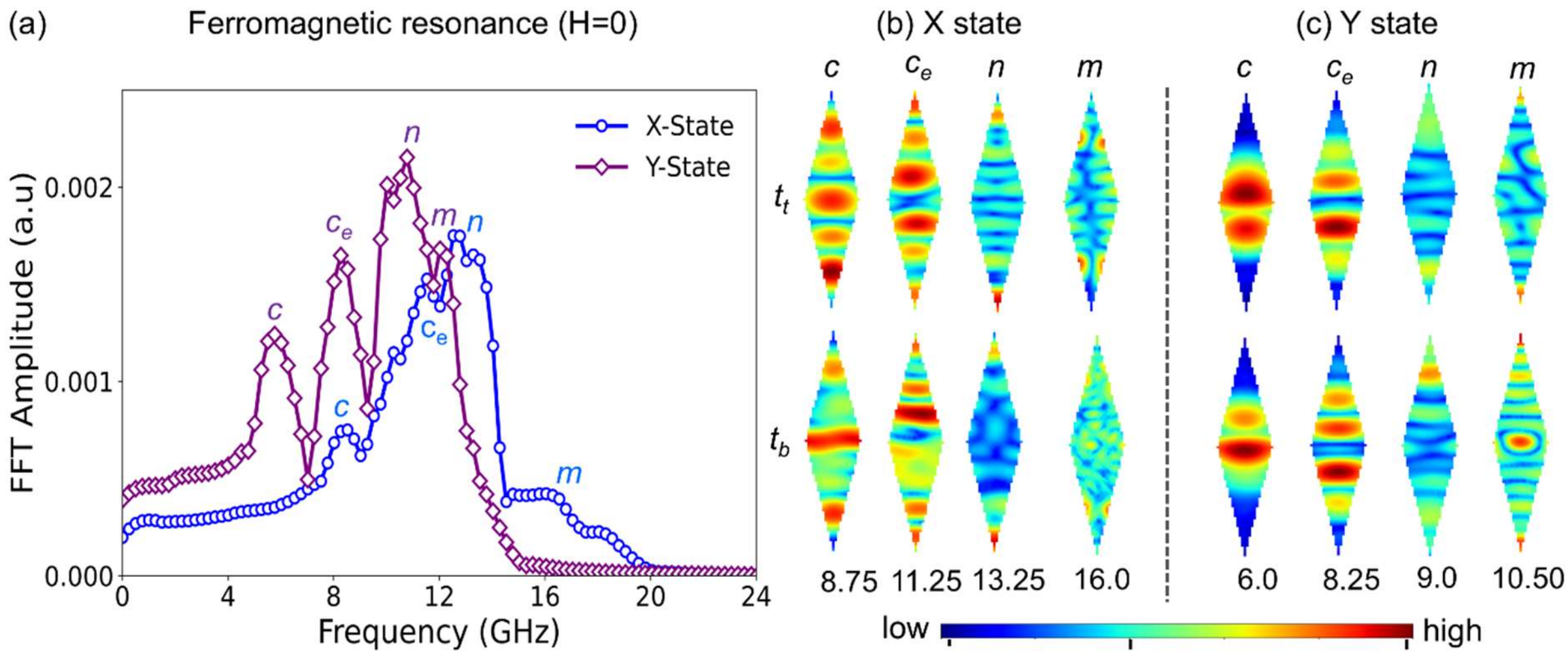

*Fig. 5.* Simulated ferromagnetic resonance (FMR) spectra and corresponding two-dimensional (2D) dynamic mode profiles of the multilayer nanomagnet (ML-1M). (a) FMR spectra for the X and Y remanent states at H = 0. (b, c) Spatial distributions of the dynamic magnetization at the indicated resonance frequencies for the X and Y states, respectively

A comparison of the two remanent states shows that changing the initialization direction modifies both the resonance frequencies and their corresponding mode profiles. In the X-state, the resonance modes appear at approximately 8.75, 11.25, 13.25, and 16.0 GHz, whereas in the Y-state, they are observed at approximately 6.0, 8.25, 9.0, and 10.50 GHz. This large frequency shift (3 GHz) originates from the different magnetic configurations established in the two remanent states. Since the two magnetic layers are aligned in anti-parallel directions in the X-state, a different internal dipolar field is produced compared with the parallel magnetization configuration in the Y-state. As a result, a distinct difference is observed in the resonance characteristics. Overall, the FMR spectra together with the corresponding 2D spatial mode profiles demonstrate that the resonance characteristics of the isolated multilayer diamond-shaped nanomagnet can be effectively controlled simply by changing the remanent magnetic state. These results suggest that the strong dipolar coupling between the magnetic layers makes this structure a promising platform for developing bias-field-free reconfigurable microwave devices.

### 3.4 Collective Spin Dynamics in Diamond-Shaped Nanomagnet Networks

To further investigate the combined effect of lateral and vertical dipolar interactions, multilayer diamond-shaped nanomagnets were arranged into three different network structures, namely ML-2M, ML-3M, and ML-4M, as shown in Fig. 6(a). In these structures, a fixed edge-to-edge separation (30 nm) was maintained between the neighbouring multilayer nanomagnets. The same lateral dimensions and layer thicknesses used for the isolated multilayer nanomagnet were also used here. By arranging additional nanomagnets, the magnetic interactions occur not only between the top and bottom magnetic layers but also between the neighbouring multilayer nanomagnets. As shown in Fig. 6(b), the magnetization reversal characteristics of these multilayer networks were investigated by simulating the hysteresis loops with the external magnetic field applied separately along the x- and y-directions. In all three networks, the magnetization changes gradually along the x-axis. However, a two-step switching process is observed along the y-axis. The intermediate plateau appearing in the y-axis hysteresis loop indicates that the magnetic layers switch sequentially at different magnetic field values rather than simultaneously. Since the hysteresis behaviour is almost identical for the ML-2M, ML-3M, and ML-4M networks, the results indicating that the dipolar interaction between the magnetic layers mainly governs the switching process.

The stable remanent magnetic configurations obtained after field initialization are shown in Fig. 6(c) and Fig. 6(d). When the structure is initialized along the x-axis, the top and bottom magnetic layers in each multilayer nanomagnet are aligned in anti-parallel directions, forming the X-state. At the same time, because of the lateral dipolar interactions, the neighbouring multilayer nanomagnets also stabilize in alternating magnetic orientations. In contrast, when the structure is initialized along the y-axis, not only the top and bottom magnetic layers but also, the neighbouring multilayer nanomagnets retain parallel magnetization, resulting in the Y-state. These results show that two different remanent magnetic states can be obtained in these multilayer networks simply by changing the initialization direction. As the number of coupled multilayer nanomagnets increases from ML-2M to ML-4M, the remanent magnetic states become more complex.

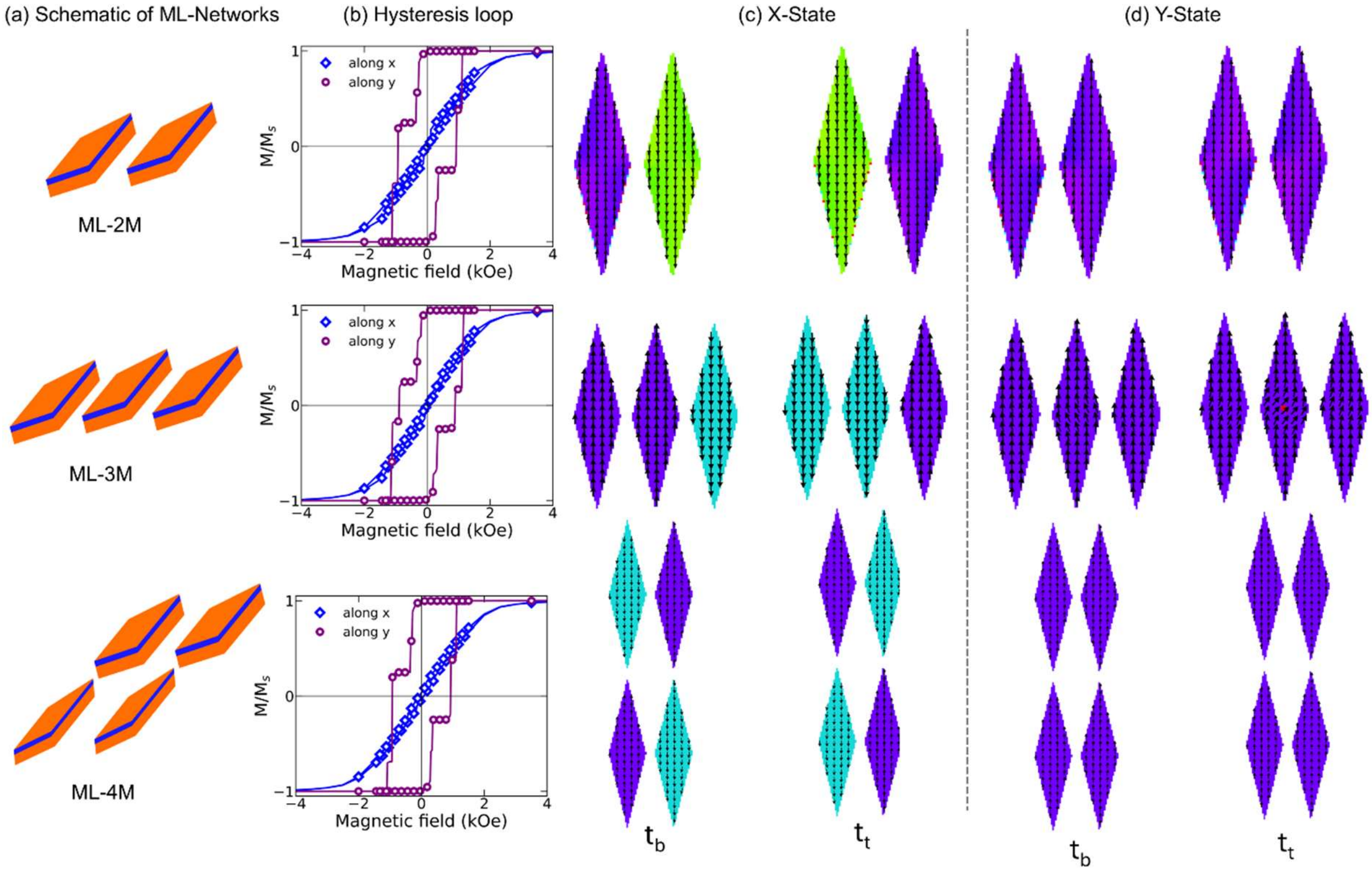

*Fig. 6.* Magnetic states and reversal behavior of coupled nanomagnet arrays. (a) Schematic of ML-2M, ML-3M, and ML-4M arrays containing two, three, and four diamond-shaped nanomagnets, respectively. (b) Magnetization hysteresis curves along the x- and y-axes. (c) Magnetization distributions at the bottom ($t_b$) and top ($t_t$) layers. (d) Corresponding magnetization patterns for the X-state at $t_b$ and $t_t$. Contrasting colors indicate opposite magnetization directions.

The simultaneous presence of vertical and lateral dipolar coupling provides greater flexibility in controlling the magnetic configurations without changing the physical geometry of the network. Such controllable magnetic states make these multilayer networks promising candidates for the development of bias-field-free reconfigurable microwave and magnonic devices.

The dynamic microwave behaviours of the dipolar-coupled multilayer diamond-shaped nanomagnet (DNM) networks were investigated by calculating the ferromagnetic resonance (FMR) spectra for the X-state and Y-state under zero external magnetic field (H = 0). The simulated FMR spectra for the three network structures are presented in Fig. 7. Multiple resonance modes are observed over a wide frequency range of approximately 4–18 GHz. The resonance frequencies and their corresponding amplitudes vary with the remanent magnetic configuration established after the field initialization process. The FMR spectra corresponding to the X-state and Y-state show distinct differences for all three network structures. In the Y-state, the dominant resonance modes are mainly distributed between 4 and 8 GHz, whereas the principal resonance modes of the X-state are shifted toward the 8–14 GHz region. This frequency variation indicates that the remanent magnetic configuration alters the internal effective magnetic field, thereby modifying the microwave response of the coupled networks. The normalized FMR spectra shown in the right panels enable a direct comparison of the resonance frequencies independent of the peak amplitudes. Several resonance modes undergo significant frequency shifts between the two remanent states, while some modes emerge or disappear after changing the initialization direction. The presence of multiple resonance modes suggests the excitation of different spin-wave modes resulting from the combined effects of lateral and vertical dipolar interactions within the multilayer networks. Although ML-2M, ML-3M, and ML-4M contain different numbers of coupled multilayer nanomagnets, they exhibit similar dynamic characteristics. The clear separation between the resonance bands of the X-state and Y-state indicates that the remanent magnetic configuration plays a key role in determining the resonance behaviours. Furthermore, the combined influence of interlayer and lateral dipolar coupling modifies the internal effective magnetic field, leading to the observed changes in the resonance frequencies.

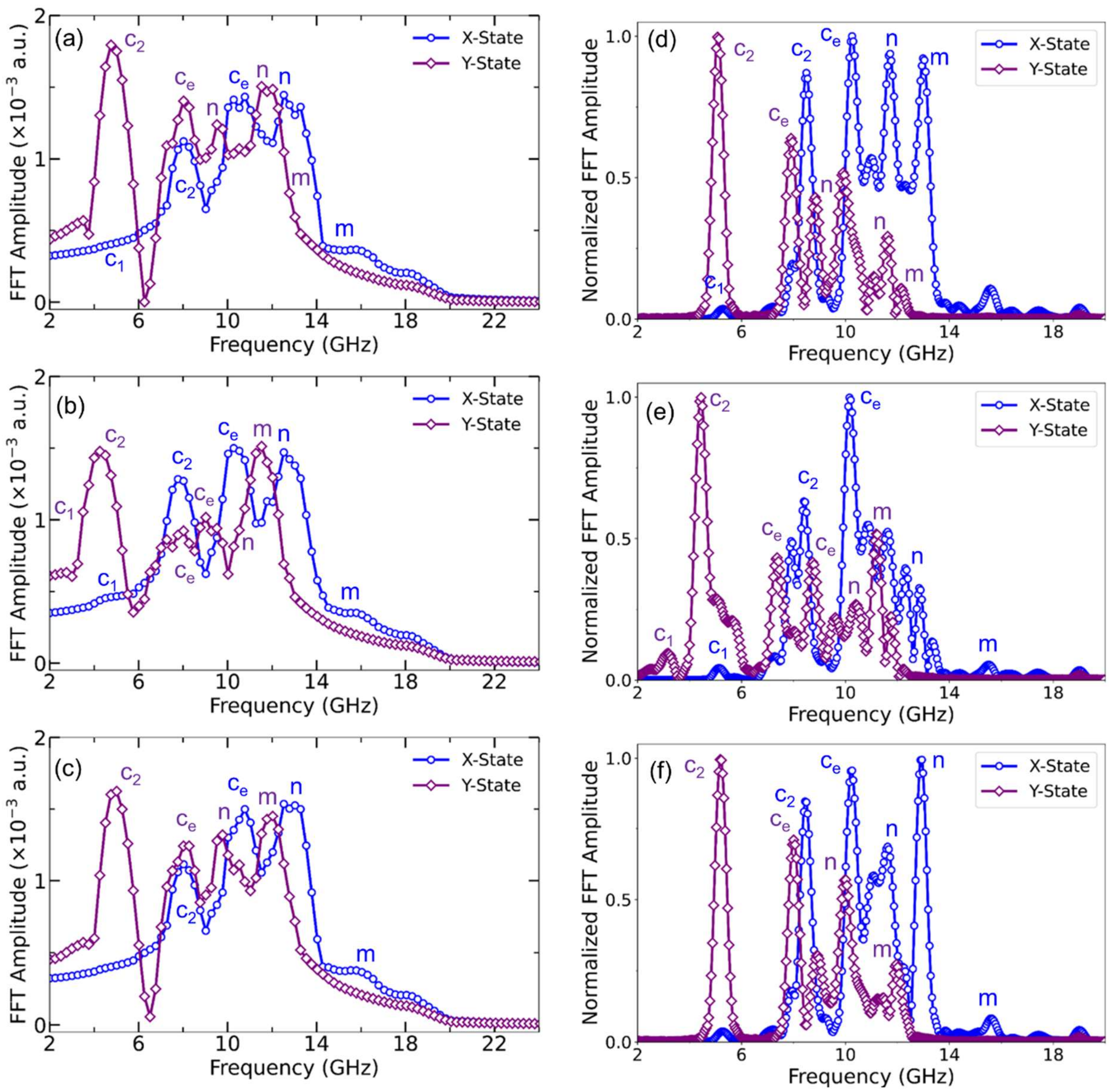

Fig. 7. Comparison of the FFT spectra obtained for the X-state and Y-state magnetic configurations. (a–c) FFT amplitude spectra of the corresponding structures, showing the resonance modes and their frequency evolution. (d–f) Corresponding normalized FFT spectra, highlighting the relative intensities and mode distributions.

Overall, these results demonstrate that the microwave response of the dipolar-coupled multilayer DNM networks can be effectively tuned by switching between two stable remanent magnetic states without applying an external bias magnetic field. This ability makes the proposed multilayer networks promising candidates for the development of bias-field-free reconfigurable microwave and magnonic devices.

**3.5 2D spatial profiles of dipolar-coupled multilayer networks.**

**3.5.1 Amplitude**

To clarify the physical origin of the resonance modes in the multilayer DNM networks, the corresponding two-dimensional (2D) spatial distributions of the dynamic magnetization were examined for both the X-state and Y-state, as shown in Fig. 8. The spatial mode profiles clearly indicate that the localization of the spin-wave excitations is strongly affected by both the remanent magnetic configuration and the number of coupled multilayer nanomagnets. According to their spatial distributions, the resonance modes can be categorized into five groups, namely $c_1$, $c_2$, $c_e$, n, and m. The $c_1$-mode is observed only in the ML-3M network and X-state of ML-2M. This mode is mainly confined to the central region of the DNM multilayer nanomagnets. It represents the lowest-order collective excitation within the coupled network. Since this mode is absent in ML-4M, its existence is closely related to the geometrical arrangement of the coupled nanomagnets. The $c_2$-mode is present in all three network structures and exhibits a relatively uniform oscillation over the central regions of the multilayer nanomagnets. Compared with the $c_1$-mode, its dynamic response extends over a larger area, indicating stronger dynamic coupling between neighbouring nanomagnets. The $c_e$-mode (mode splitting) is also observed in every network configuration and occupies a broader spatial region than the $c_2$-mode. The dynamic magnetization extends over a significant portion of both the top ($t_t$) and bottom ($t_b$) magnetic layers, indicating that this mode originates from collective oscillations driven by the combined influence of vertical and lateral dipolar interactions. The nearly symmetric distributions observed in several structures further suggest efficient coupling between neighbouring multilayer nanomagnets. The n-mode is distinguished by the appearance of one or more nodal regions, where the oscillation amplitude becomes very small. These nodal regions divide the nanomagnets into separate oscillating sections, confirming that the n-mode corresponds to a higher-order standing spin-wave excitation. As the network size increases from ML-2M to ML-4M, the nodal pattern becomes more prominent because of the increasingly complex dipolar field distribution within the coupled system.

The m-mode corresponds to the highest-order resonance observed in these networks. It exhibits highly non-uniform magnetization with several localized regions of high and low oscillation amplitude distributed throughout the multilayer nanomagnets. This complicated spatial pattern is attributed to strong mode hybridization produced by the combined effects of interlayer coupling and interactions between neighbouring nanomagnets. Among all the identified modes, the m-mode possesses the most complex spatial distribution and generally appears at the highest resonance frequencies. A comparison between the X-state and Y-state further shows that changing the remanent magnetic configuration modifies not only the resonance frequencies but also the spatial characteristics of the spin-wave modes as shown in Fig. 8(a & b). Although some modes retain similar overall features in both magnetic states, others exhibit noticeable changes in localization, nodal structure, and oscillation amplitude. In particular, the Y-state generally displays more symmetric mode distributions because the magnetic moments of the multilayer nanomagnets remain predominantly parallel. In contrast, the X-state exhibits more complicated spatial patterns owing to the antiparallel alignment between the top and bottom magnetic layers. This alignment changes the internal dipolar field distribution and consequently alters the dynamic magnetization pattern. Furthermore, as the number of coupled multilayer nanomagnets increases from ML-2M to ML-4M, the mode profiles become progressively more complex. The presence of additional neighbouring nanomagnets strengthens the lateral dipolar interaction, leading to enhanced collective oscillations, stronger mode localization, and more pronounced standing spin-wave characteristics. These observations demonstrate that the spatial characteristics of the resonance modes can be effectively tailored by controlling both the network geometry and the remanent magnetic configuration.

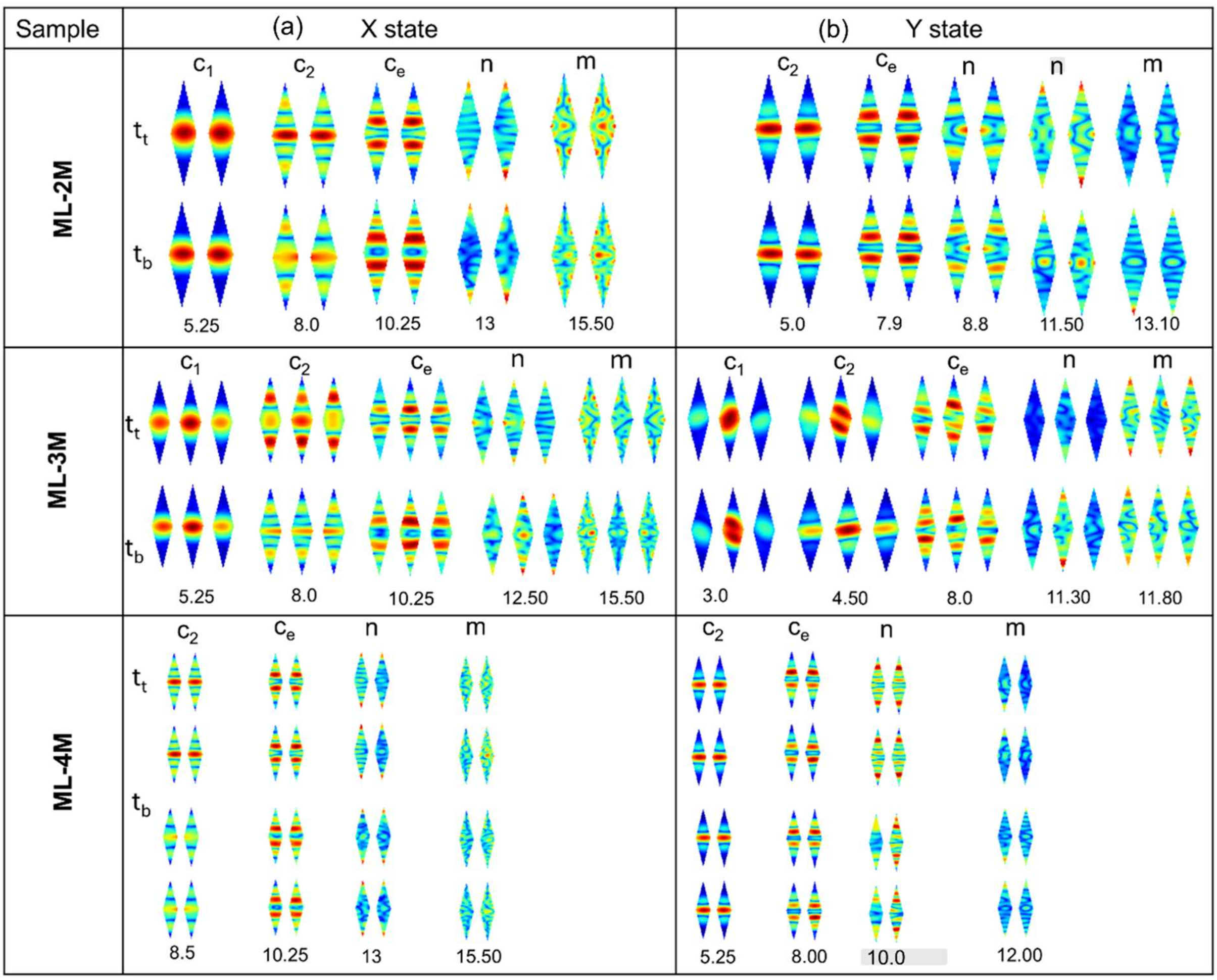

Fig. 8. Spatial profiles of FMR modes in the (a) X-state and (b) Y-state for the ML-2M, ML-3M, and ML-4M structures. The corresponding mode frequencies are indicated below each 2D-profile.

Overall, the 2D spatial mode profiles clearly demonstrate that the observed resonance modes originate from different dynamic magnetization distributions governed by the combined effects of vertical interlayer coupling, lateral dipolar interactions, and the remanent magnetic state. These findings confirm that the proposed multilayer DNM networks support multiple controllable spin-wave modes, highlighting their potential for the realization of bias-field-free reconfigurable microwave and magnonic devices.

Further studies will focus on gaining deeper insight into the characteristics of the observed modes and the mechanisms responsible for their coupling and origin of the shift. Attention will be given to the spatial distribution of the modes, possible mode hybridization (phase and phase angle), ultrafast switching and estimating the energy per bit.

## VI. SUMMARY

The microwave properties of a diamond-shaped nanomagnet (DNM) can be modified as required without the need for an external bias magnetic field. The static and dynamic magnetic properties of the DNM were investigated using micromagnetic simulations. Magnetization reversal, switching between different magnetic states, and the corresponding ferromagnetic resonance (FMR) characteristics were primarily examined. Two stable magnetic states, referred to as the X-state and Y-state, were obtained through a simple magnetic-field initialization procedure. These two states exhibit distinct FMR spectra. Therefore, switching between the X-state and Y-state enables modification of the microwave response.

Overall, the results demonstrate that the magnetic state of the diamond-shaped nanomagnet, along with its corresponding FMR and microwave properties, can be controlled without an external bias magnetic field. In addition, the different dynamic modes can be characterized through analyses of their amplitude and phase profiles. These findings suggest that the proposed DNM could serve as a potential platform for the future development of reconfigurable microwave and magnonic devices, as well as frequency-tunable microwave devices.

## ACKNOWLEDGMENTS

KB acknowledges the National Institute of Technology Andhra Pradesh, India, for providing the opportunity to hold an Ad Hoc Faculty position and pursue teaching and research activities.